# The Swedish System of Innovation:
# Regional Synergies in a Knowledge-Based Economy

Loet Leydesdorff [a] & Øivind Strand [b]

**Abstract**

Based on the complete set of firm data for Sweden ($N = 1,187,421$; November 2011), we analyze the mutual information among the geographical, technological, and organizational distributions in terms of synergies at regional and national levels. Mutual information in three dimensions can become negative and thus indicate a net export of uncertainty by a system or, in other words, synergy in how knowledge functions are distributed over the carriers. Aggregation at the regional level (NUTS3) of the data organized at the municipal level (NUTS5) shows that 48.5% of the regional synergy is provided by the three metropolitan regions of Stockholm, Gothenburg, and Malmö/Lund. Sweden can be considered as a centralized and hierarchically organized system. Our results accord with other statistics, but this Triple Helix indicator measures synergy more specifically and quantitatively. The analysis also provides us with validation for using this measure in previous studies of more regionalized systems of innovation (such as Hungary and Norway).

**Keywords**

knowledge base, measurement, Sweden, Triple Helix, indicator, regions, national system

---

[a] University of Amsterdam, Amsterdam School of Communication Research (ASCoR), Kloveniersburgwal 48, 1012 CX Amsterdam, The Netherlands; loet@leydesdorff.net
[b] Aalesund University College, Department of International Marketing, PO Box 1517, 6025 Aalesund, Norway; +47 70 16 12 00; ost@hials.no .

**Introduction**

The metaphor of a Triple Helix of University-Industry-Government Relations emerged as an operationalization of the complex dynamics of innovation during the second half of the 1990s (Etzkowitz & Leydesdorff, 1995 and 2000). Etzkowitz (1994) first contributed with a chapter about university-industry relations to a volume entitled *Evolutionary Economics and Chaos Theory: New directions for technology studies* (Leydesdorff & Van den Besselaar, 1994). At the time, the notion of a virtual hyper-cycle and its possible effects in terms of lock-ins, path-dependencies, and bifurcation were central to the discussion (e.g., Arthur, 1989; Bruckner *et al*., 1994; David, 1988; David & Foray, 1994), but yet poorly elaborated—beyond the formal modeling—into theories relevant for science, technology, and innovation studies. In evolutionary economics, a focus on complex systems had been introduced in relation to the study of "national *systems* of innovation" (Freeman, 1987; Lundvall, 1988, 1992; Nelson, 1993; cf. Carlson, 2006; Utterback & Suarez, 1993).

In the Epilogue to this volume, Leydesdorff (1994, at pp. 186f.) formulated that a system with three subdynamics would be sufficiently complex to encompass the various species of chaotic behavior (Li & Yorke, 1975). Accordingly, he suggested distinguishing a dynamics of regulation (by governments) in addition to Schumpeter's well-known distinction between economic factor substitution and technological innovation (Schumpeter, 1939; Sahal, 1981; cf. Nelson & Winter, 1977, 1982). When the three (sub)dynamics are spanned orthogonally, this can be elaborated as in Figure 1.

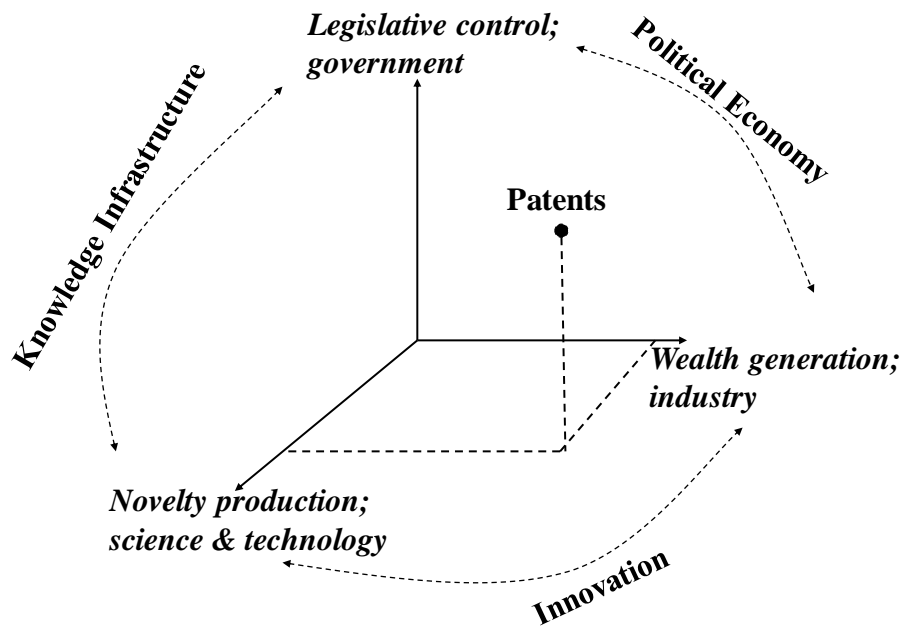

**Figure 1**: Patents as events in the three-dimensional space of Triple Helix interactions. (Source: Leydesdorff, 2010a: 370).

Patents, for example, can be considered as outputs of the science system, but inputs into the economy; at the same time, patents function in the legal regime of intellectual property protection. Whereas two subdynamics can be expected to shape each other in coevolutions and thus lead to trajectories, one additional selection environment can move observable trajectories into virtual regimes (Leydesdorff & Van den Besselaar, 1998). The extension of the study of university-industry relations (Etzkowitz, 2002) to the study of university-industry-government relations might thus enable us to operationalize the relation between the historical phenomena—as studied in case studies—and the evolutionary dynamics. In innovation policies, one is interested in possibilities "adjacent" to those that have actually occurred historically (Kauffman, 2000).

In the further elaboration of the Triple Helix model, Etzkowitz & Leydesdorff (2000) proposed considering the emerging overlay of communications in relations of university, industry, and government as a potential hyper-cycle of meaning exchange processes which can operate as a feedback or feedforward on the three underlying spheres of communication, and thus shape new opportunities in frictions between institutional missions and functional competencies. How could one move this theoretical heuristics towards measurement and modeling? Leydesdorff (2003) turned to Ulanowicz' (1986: 143) proposal to model these virtual mechanisms by using mutual information in three dimensions.

Unlike Shannon-type information—that is, uncertainty—mutual information in three (or more) dimensions can be positive or negative (or zero) indicating an increase or decrease of uncertainty, respectively. Yeung (2008: 59f.) proposed calling this information (denoted by him as $\mu$) a *signed* information measure. Krippendorff (2009a) showed that this signed measure is incompatible with Shannon's linear notion of communication, but can be considered as the difference between (Shannon-type) information ($I$) necessarily generated in complex interactions (according to the second law)[1] versus redundancy ($R$) generated in loops of communication in which the information is differently provided with meanings (Krippendorff, 2009b): $\mu = I - R$ (Leydesdorff, 2010b, 2011). A value of $\mu < 0$ indicates that more redundancy is generated in a system than information produced, and under this condition uncertainty is reduced at the level of the system. In other words, the system thus contains negentropy (Brillouin, 1962), and this reduction of uncertainty can be specified given the various uncertainties in distributions and their interactions in a configuration.

In a series of studies, we explored this measure initially using institutional addresses in publications downloaded from the Science Citation Index (SCI) at the Web of Science (WoS) of

---

[1] Thermodynamic entropy $S = k_B * H$ since according to Gibbs' formula of entropy: $S = -k_B \sum_i p_i \log(p_i)$. While $k_B$ is a constant (the Boltzmann constant), the second law of thermodynamics is equally valid for probabilistic entropy. The dimensionality of $H$, however, is not defined by the formula, since both $k_B$ and $S$ are expressed as Joule/Kelvin.

Thomson Reuters. Leydesdorff (2003), for example, used a string of searches in order to distinguish between academic, industrial, or governmental addresses. Park *et al*. (2005) used this same string for international comparisons among countries and world regions (Ye *et al*., in preparation). In the meantime, Sun *et al*. (2007) generated a complete database with precise author-addresses in Japan. This data allowed us to test the model with greater precision (Leydesdorff & Sun, 2009). Among other things, these authors concluded that the international dimension in coauthoring counter-acted on the local integration among authors with different institutional affiliations at the national level of Japan.

In the case of the Japanese science system, international coauthorship relations function as a subdynamics of the national system, since the system itself tends to erode—that is, to generate more entropy than redundancy—without these international relations. Using similar measures, Kwon *et al*. (2010) found much more stable patterns for Korea, and Ye *et al*. (in preparation) studied how countries have differed since the second half of the 1990s in terms of incorporating these effects of globalization. The Triple Helix thesis, which is dated 1995, may thus be in need of an update (Lawton-Smith & Leydesdorff, in preparation). A local configuration can also be considered as meta-stable, that is, potentially globalizing (Leydesdorff & Deakin, 2011). In summary, these studies taught us that in addition to local integration in university-industry-government relations, one should also account for the dimension globalization-localization or, in other words, the international dimension as important to further development at the national systems level.

In a parallel series of studies, we used firms as units of analysis for studying the knowledge-based economy (Leydesdorff, 2006, 2010a). Whereas publications may have three (or more) types of addresses in dimensions that can co-vary to different extents, firms can be attributed similarly with three (or more) analytically independent dimensions such as geographical addresses, organizational size, and technological core competencies, etc. Since data with three or more independent dimensions (e.g., geographical addresses, OECD classifications of technology, and organizational size) was available for the Netherlands and Germany at respectively the level of individual firms and municipalities, maps could be made of these countries indicating where synergies were generated at regional levels (Leydesdorff & Fritsch, 2006; Leydesdorff *et al*., 2006).

In the case of Germany, we found reduction of uncertainty—here considered as a proxy for synergy among the functions—at the level of federal states (*Länder*), while in the Netherlands surplus value was found at the national level. Furthermore, the synergy at the level of *Länder* was at the time different for states which had previously belonged to the East-German GDR and for the more knowledge-based states in the West, whereas at a lower level of aggregation (*Regierungsbezirke*) this historical condition faded away during the 1990s. Furthermore, the "embeddedness" of medium-tech manufacturing was found to be more important for generating synergy than high-tech (Cohen & Levinthal, 1990); knowledge-intensive services are not constrained regionally and can thus be expected to contribute to the delocalization ("footlooseness"; cf. Vernon, 1979) of the economy.

In a follow-up study for Hungary, but using a less complete dataset, Lengyel & Leydesdorff (2011) concluded that Hungary as a nation no longer provides surplus synergy to three regional subsystems. This disintegration of the "national" system followed during the 1990s after the demise of the Soviet Union, when the country first went into "transition" and then became an accession country to the EU. The most important subsystem in Hungary was found in the metropolitan area of Budapest and its environment. The western parts of the country were no longer integrated nationally, but indeed had "access" to relevant environments in Austria, Germany, and the remainder of the EU, whereas some eastern provinces remained more under the regime of the old (state-controlled) system. Although these conclusions were tentative, they could be supported by another reading of existing statistics.

In a similar vein, Strand & Leydesdorff (in press) concluded on the basis of the full set of Norwegian firm data that the knowledge-based system of this national system was no longer integrated in relation to the national universities, but driven by foreign investments in offshore industries in the western parts of the country. Internationalization and globalization in the latter two studies seemed thus core dimensions for understanding how the knowledge-based economy operates. The speculative character of some of these conclusions, however, made us turn to the Swedish innovation system for comparison.

More than any other, the Swedish innovation system has been studied in detail, is documented statistically with great precision, and can thus provide us with a benchmark to test our Triple Helix methodology. In this study, we use the full set of data about 1,187,421 firms made available by Statistics Sweden (November 2011), and analyze this data with a design similar to the previous studies. The Swedish case, however, allows us to specify an expectation.

**The knowledge base of the Swedish economy**

The Swedish economy can be considered as a mixed economy of free-market activities and government interventions. Although liberalization in recent decades has affected the system, the welfare state model is still prevalent, perhaps more than in any other European country. One can thus expect a national system of innovation with surplus value at the national level over and above the sum of the regional innovation systems. The modern nation state has further been developed since Napoleonic times, reinforcing a national identity that has been shaped since the Reformation.

Sweden spends 3.42% of its GDP on R&D (2010 data; Eurostat, 2010) and ranks in this respect number two—after Finland with 3.87%—among the European nations. The knowledge infrastructure of Sweden is centralized in three metropolitan areas: Stockholm, Gothenburg, and Malmö. The capital region of Stockholm hosts Karolinska Institute—one of Europe's largest medical universities—Stockholm University, the Royal Institute of Technology, and the Stockholm School of Economics. Uppsala University (founded in 1477) is located in the neighboring county of Stockholm. Gothenburg, located in Västre Göthaland, hosts the University of Gothenburg and Chalmers University of Technology. The largest university, Lund University, is located in Skåne (that is, the region surrounding Malmö).

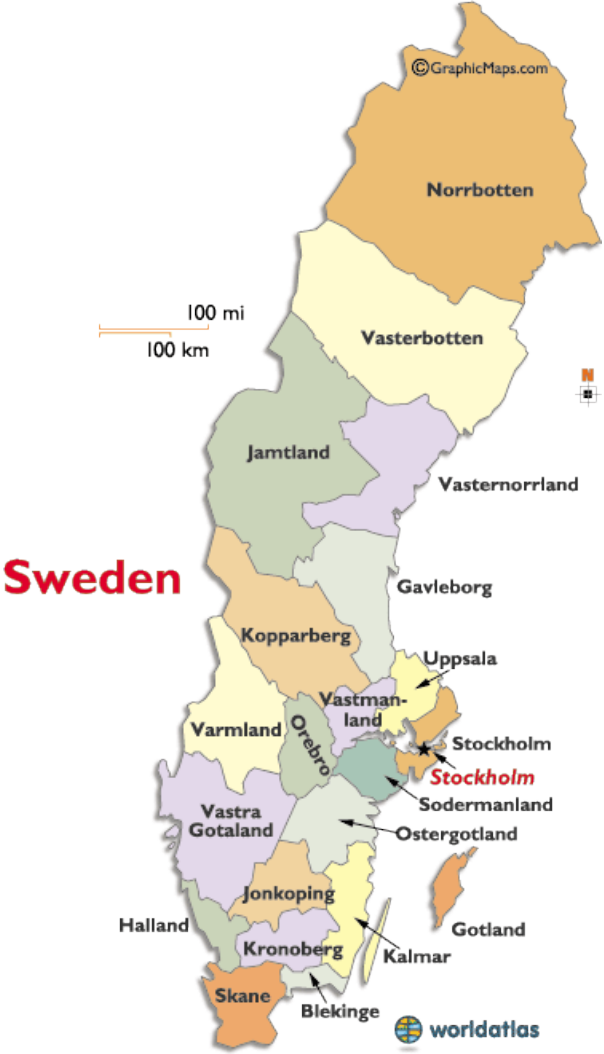

**Figure 2**: Location of the 21 Swedish counties at the NUTS 3 level.[2] (Source: http://www.worldatlas.com/webimage/countrys/europe/lgcolor/secounties.htm.)

Sweden is organized at three levels of government: the central government (NUTS 0 level),[2] 21 counties (at the NUTS3 level; Figure 2) and 290 municipalities (at the NUTS5 level). Benner and Sandström (2000) argued that institutionalization of a Triple Helix is critically dependent upon research funding. Table 1 lists the distribution of R&D expenditures over the 21 counties. This distribution is very skewed: the per capita R&D expenditure ranges from US$2,970 in Östergötaland to US$54 in Jämtland.[3] The percentage budget distribution of R&D funding over the regions is also extremely skewed: the shares among counties range from more than 33% in Stockholm and 21.7% in Västre Götalands (Gothenburg) to below 0.1% for some other counties (Table 1). (The distribution of firms over counties is provided in Table 4.)

**Table 1**: Geographical distribution of R&D funding and firms in Sweden over 21 counties (2009 data). (Source : Statistics Sweden, 2011c.)

| County[4] | R&D Expenditure Per Capita 2009 SEK[3] (a) | R&D Expenditure 2009 in million SEK (b) | Idem as a percentage of the national GERD[5] (c) |
|---|---|---|---|
| 01 Stockholm county | 18,306 | 36,882 | 33.0 |
| 03 Uppsala county | 19,999 | 6,629 | 5.9 |
| 04 Södermanland county | 4,646 | 1,248 | 1.2 |
| 05 Östergötland county | 20,571 | 8,780 | 7.9 |
| 06 Jönköping county | 4,656 | 1,564 | 1.4 |
| 07 Kronoberg county | 5,123 | 938 | 0.8 |
| 08 Kalmar county | 2,288 | 535 | 0.5 |
| 09 Gotland county | 699 | 40 | 0.0 |
| 10 Blekinge county | 8,291 | 1,266 | 1.1 |
| 12 Skåne county | 13,893 | 17,078 | 15.3 |
| 13 Halland county | 2,047 | 607 | 0.5 |

---

[2] NUTS is an abbreviation for "*Nomenclature des Unités Territoriales Statistiques*" (that is, Nomenclature of Territorial Units for Statistics). The NUTS classification is a hierarchical system for dividing up the economic territory of the EU.
[3] Swedish Kroner; SEK 1 is approximately equivalent to US$ 0.14.
[4] We follow the administrative numbering of the 21 counties from 1 to 25 by Statistics Sweden (2011c).
[5] Gross expenditure in R&D (GERD) is defined by OECD (2012).

| | | | |
|---|---|---|---|
| 14 Västra Götaland county | 15,424 | 24,191 | 21.7 |
| 17 Värmland county | 2,129 | 582 | 0.5 |
| 18 Örebro county | 5,970 | 1,664 | 1.5 |
| 19 Västmanland county | 5,856 | 1,470 | 1.3 |
| 20 Dalarna county | 2,102 | 581 | 0.5 |
| 21 Gävleborg county | 4,901 | 1,354 | 1.2 |
| 22 Västernorrland county | 3,978 | 967 | 0.9 |
| 23 Jämtland county | 371 | 47 | 0.0 |
| 24 Västerbotten county | 11,112 | 2,871 | 2.6 |
| 25 Norrbotten county | 5,281 | 1,316 | 1.2 |

The Swedish innovation system and its components have been thoroughly analysed by a large number of scholars. Most of these studies are based on (sometimes comparative) case-study research of industrial sectors and/or technologies; for example, the biotechnology sector in Sweden was studied by Cooke (2005), Rosiello (2007), Brink *et al.*, (2007), and Modyson *et al.* (2008). The fashion industry was studied by Hauge *et al.* (2009), and the music industry by Power & Hallencreutzer (2005). The roles of various universities have been studied by Jones-Evans & Klofsten (1997). Berggren & Lindholm Dahlstrand (2009) and Benneworth *et al.* (2009) focused on Lund University.

A number of studies have also addressed regional cluster formation (e.g., Lundequist & Power, 2002; Hallencreutzer & Lundequist, 2003). Using quantitative methods, the knowledge base of regional innovation systems was studied by Asheim & Gertler (2005); Martin (2012) used register data for this purpose. The knowledge bases of various innovation systems in Scania have been studied by Martin & Moodyson (2011), whereas Asheim & Conen (2005) studied the knowledge base of regional innovation systems in the three Nordic countries (Norway, Sweden, and Denmark; cf. Fagerberg *et al.*, 2009a and b).

In summary, the Swedish innovation system has been studied in considerable detail. In our opinion, however, the excellent data collected by Statistics Sweden has been under-exploited, since most studies collected their own specific data, which makes it difficult to compare across studies and among regions and/or sectors of industry. In the present study, we take a top-down approach subjecting all available data to statistical decomposition analysis (Theil, 1972). Statistical decomposition analysis takes full advantage of the aggregation in Shannon's (1948) entropy statistics, and allows us to specify between-group uncertainty and within-group uncertainty precisely in bits of information.

**Methods and materials**

*a.    Data*

The data was received from Statistics Sweden in December 2011 following collection in November 2011. The collection is comprehensive since firms are legally obliged to respond to the questions of Statistics Sweden; the full set contains 1,187,421 Swedish firms at the time. The records made available to the public contain tables with variables specified at the level of municipalities. Among these variables are three variables that we can use as independent operationalizations of the three dimensions of the Triple Helix: the technological classifications of firms by the OECD, the geographical classification of Eurostat that corresponds to local, regional, and national government structures, and we use firm-size as an operationalization of the economic scale of a firm's operations.

Geographical address information is available in the form of the four-digit code of the (290) municipalities; this data can be aggregated straightforwardly to counties (*Län*), regions (*Riksområden*), lands (*Landsdelar*), and the nation (in this case, Sweden). The municipality is the lowest geographical level of analysis (NUTS5) and the lowest level of administration in Sweden. The 21 counties provide the second level of administration in Sweden (at the NUTS3 level). The regional and national levels (NUTS2 and NUTS1, respectively) will also be included in our analysis for the sake of comparison with OECD and Eurostat statistics, but NUTS2 is not used as a separate level of administration in Sweden.

The Swedish data is finer-grained than in our previous studies. There are 290 units at the lowest (NUTS5) level of municipalities, whereas the Hungarian data contained 168 sub-regions, the Dutch set contained 90 postcodes, and the German study was based on data for 438 NUTS3 regions. Technology is indicated in our data by means of the one-letter sector classification used by Statistics Sweden. The code is listed in Appendix A. Unfortunately, this code is less specific than the NACE codes of the OECD.[6] The organizational dimension is indicated by company size in terms of the number of employees (as in previous studies). The classes and their values are tabulated in Table 2.

---

[6] NACE stands for Nomenclature générale des Activités économiques dans les Communautés Européennes. The NACE code can be translated into the International Standard Industrial Classificiation (ISIC). A concordance table between the Swedish sector classification and the NACE codes can be found at
http://www.scb.se/Grupp/Hitta_statistik/Forsta_Statistik/Klassifikationer/_Dokument/070129kortversionSnisorterad 2007.pdf .

**Table 2**: Distribution of employees in the Swedish data and corresponding uncertainty. (Source: Statistics Sweden, 2011b).

| Size | Number of employees | Number of companies | Probability | Uncertainty in bits |
|---|---|---|---|---|
| 1 | 0 | 832,840 | 0.7014 | 0.3589 |
| 2 | 1-4 | 219,033 | 0.1845 | 0.4498 |
| 3 | 5-9 | 59,615 | 0.0502 | 0.2169 |
| 4 | 10-19 | 37,532 | 0.0316 | 0.1575 |
| 5 | 20-49 | 24,496 | 0.0206 | 0.1155 |
| 6 | 50-99 | 8,520 | 0.0072 | 0.0511 |
| 7 | 100-199 | 3,458 | 0.0029 | 0.0245 |
| 8 | 200-499 | 1,413 | 0.0012 | 0.0116 |
| 9 | >500 | 514 | 0.0004 | 0.0048 |
|   |   | 1,187,421 | 1.0000 | 1.3905 |

The size of a company can be considered as a proxy of its organization and economic dynamics (e.g., Pugh *et al.*, 1969a and b; Blau and Schoenherr, 1971). For example, small and medium-sized companies can be expected to operate differently from large multinational corporations. We used this proxy throughout our series of studies, hitherto. Unlike the number of researchers or the level of R&D spending—which is sometimes classified information—this size of the company in terms of the number of employees is available in public databases. We use this indicator also because one can expect an R&D-indicator to be correlated with the NACE codes which we use as the technology indicator in terms of being R&D-intensive or not; the two indicators would then partially reflect the same dimension whereas the organizational size can be considered as an independent (and economically determined) dimension.

Companies without employees account for more than 70% of the companies in Sweden, unlike the Hungarian (29.8%), and Dutch studies (19.7%). In the German study, neither this class of companies nor the number of self-employed in firms were included because of the different way this data was collected.

*b.    Methods*

As noted above, the mutual information in more than three dimensions—the Triple-Helix indicator to be used here—is a signed information measure (Yeung, 2008), and therefore not a Shannon-information (Krippendorff, 2009). However, this measure is derived in the context of information theory and follows from the Shannon formulas (e.g., Abramson, 1963; Ashby, 1964; McGill, 1954).

According to Shannon (1948) the uncertainty in the relative frequency distribution of a random variable $x$ ($\sum_x p_x$) can be defined as $H_X = -\sum_x p_x \log_2 p_x$. Shannon denotes this as probabilistic entropy, which is expressed in bits of information if the number two is used as the base for the logarithm. (When multiplied by the Boltzman constant $k_B$, one obtains thermodynamic entropy and the corresponding dimensionality in Joule/Kelvin. Unlike thermodynamic entropy, probabilistic entropy is dimensionless and therefore yet to be provided with meaning when a system of reference is specified.)

Likewise, uncertainty in a two-dimensional probability distribution can be defined as $H_{XY} = -\sum_x \sum_y p_{xy} \log_2 p_{xy}$. In the case of interaction between the two dimensions, the uncertainty is reduced with the mutual information or transmission: $T_{XY} = (H_X + H_Y) - H_{XY}$. If the distributions are completely independent $H_{XY} = H_X + H_Y$, and consequently $T_{XY} = 0$.

In the case of three potentially interacting dimensions ($x$, $y$, and $z$), the mutual information can be derived (e.g., Abramson, 1963: 131 ff.):

$$T_{XYZ} = H_X + H_Y + H_Z - H_{XY} - H_{XZ} - H_{YZ} + H_{XYZ} \qquad (1)$$

The interpretation is as follows: association information can be categorized broadly into correlation information and interaction information. The correlation information among the attributes of a data set can be interpreted as the total amount of information shared between the attributes. The interaction information can be interpreted as multivariate dependencies among the attributes. A spurious correlation in a third attribute, for example, can reduce the uncertainty between the other two.

Compared with correlation, mutual information can be considered as a more parsimonious measure for the association. The multivariate extension to mutual information was first introduced by McGill (1954) as a generalization of Shannon's mutual information. The measure is similar to the analysis of variance, but uncertainty analysis remains more abstract and does not require assumptions about the metric properties of the variables (Garner & McGill, 1956). Han (1980) further developed the concept; positive and negative interactions were also discussed by Tsujishita (1995), Jakulin (2005), and Yeung (2008: 59 ff.).

Our calculations contain three single-parameter uncertainties: geographical ($H_G$), technological ($H_T$), and organizational uncertainty ($H_O$). The three two-parameter uncertainties are: $H_{GT}$, $H_{GO}$, and $H_{TO}$; the three-parameter uncertainty is denoted $H_{GTO}$. Similarly, the calculations contain three two-parameter transmissions ($T_{GT}$, $T_{GO}$, $T_{TO}$) and one three-parameter transmission $T_{GTO}$. The numerical results, however, are abstract and not yet meaningful; they need to be appreciated by means of substantive theorizing. As noted, we appreciate the values of the transmissions as indicators of interactions among the three knowledge functions specified above that may lead to synergy in one configuration more than in another. We enrich the discussion further with other concepts, but one should be aware that our appreciation has mainly the status of providing heuristics for raising questions for further research suggested by our results.

One of the advantages of entropy statistics is that the values can be fully decomposed. Analogously to the decomposition of probabilistic entropy (Theil, 1972: 20f.), the mutual information in three dimensions can be decomposed into groups as follows:

$$T = T_0 + \sum_G \frac{n_G}{N} T_G \tag{2}$$

Since we will decompose in the geographical dimension, $T_0$ is between-region uncertainty; $T_G$ is the uncertainty prevailing at a geographical scale $G$; $n_G$ is the number of firms at this geographical scale; and $N$ the total number of firms in the whole set ($N = 1{,}187{,}421$).

The between-group uncertainty ($T_0$) can be considered as a measure of the dividedness.[7] A negative value of $T_0$ indicates an additional synergy at the higher level of national

---

[7] $T_0$ between two parameters can also be negative because $T_{xy} = H_x + H_y - H_{xy}$ and the decomposition consequently is:

$$T_0^{xy} + \sum_i \frac{n_{xy}}{N} T_{xy} = (H_0^x + \sum_i \frac{n_x}{N} H_x) + (H_0^y + \sum_i \frac{n_y}{N} H_y) - (H_0^{xy} + \sum_i \frac{n_{xy}}{N} H_{xy}) \tag{3}$$

Or rewritten:

$$T_0^{xy} + \sum_i \frac{n_{xy}}{N} T_{xy} = (H_0^x + H_0^y - H_0^{xy}) + (\sum_i \frac{n_x}{N} H_x + \sum_i \frac{n_y}{N} H_y - \sum_i \frac{n_{xy}}{N} H_{xy}) \tag{4}$$

The right-hand term $\sum_i \frac{n_{xy}}{N} T_{xy} > 0$ because each term in the summation is necessarily positive (Shannon, 1948; Theil, 1972). The left-hand term, however, can be negative if $H_0^x + H_0^y < H_0^{xy}$ because uncertainty can vary among regions, but differently in the dimensions $x$ and $y$, or mutual information ($T_{xy}$) between them.

agglomeration among the lower-level geographical units. In the Netherlands and Norway, for example, a surplus was found at the national level; in Germany, this surplus was found at the level of the federal states (*Länder*). Note that one cannot compare the quantitative values of $T_0$ across countries—because these values are sample-specific—but one is allowed to compare the dividedness in terms of the positive or negative signs of $T_0$.

**Results**

*a. NUTS3 level* ("*Län*")
Table 3 provides the uncertainties for the 21 counties (*Län*) of Sweden, expressed as percentages of the maximum entropy in the corresponding category. Uncertainties in the distribution were normalized as a percentage of maximum entropy given the number of municipalities in each county (NUTS5). Analogously, uncertainty in the technology dimension is scaled to the maximum entropy given the number of industry sectors (that is, $\log_2 22$), and uncertainty in the organizational proxy is normalized with reference to the maximum number of classes (that is, $\log_2 9$). A scaled uncertainty of 100% in the geographical distribution at county level, for example, would indicate that firms are equally distributed among the municipalities in a county.

---

One interpretation is that the next-order level adds to the uncertainty in each of the two dimensions less than to their joint entropy, and thus the joint entropy is relatively larger in the regions. In our case (of a nation and regions), a negative value of $T_0$ in a bilateral coupling (e.g., $T_{GO}$ or $T_{GT}$) means that the mutual information in the regions is lower on average than at the national level. (Less coupling leads to more joint entropy.) In the case of $T_0 < 0$, the national level reduces uncertainty relative to the sum of the regions, and thus operates as a *system* with reference to this mutual information. Note that the between-group uncertainty can be negative in one interaction and positive in another.

**Table 3**: Normalized results for uncertainties at county level (NUTS3), expressed as percentages of the maximum uncertainty.

| | $H_G/$ max($H_G$) (a) | $H_O/$ max($H_O$) (b) | $H_T/$ max($H_T$) (c) | $H_{GO}/$ max($H_{GO}$) (d) | $H_{GT}/$ max($H_{GT}$) (e) | $H_{TO}/$ max($H_{TO}$) (f) | $H_{GTO}/$ Max($H_{GTO}$) (g) |
|---|---|---|---|---|---|---|---|
| *01 Stockholms län* | 67,9 | 45,1 | 82,3 | 58,6 | 74,2 | 65,6 | 65,8 |
| *03 Uppsala län* | 71,0 | 42,0 | 82,6 | 56,0 | 76,9 | 64,0 | 65,1 |
| *04 Södermanlands län* | 87,9 | 46,2 | 82,9 | 66,8 | 84,6 | 65,7 | 71,7 |
| *05 Östergötlands län* | 78,3 | 47,6 | 82,5 | 64,0 | 79,8 | 65,9 | 69,1 |
| *06 Jönköpings län* | 87,0 | 44,6 | 75,5 | 67,3 | 79,9 | 59,4 | 67,6 |
| *07 Kronobergs län* | 86,1 | 39,3 | 72,5 | 61,9 | 77,3 | 55,7 | 63,6 |
| *08 Kalmar län* | 93,5 | 42,8 | 76,1 | 69,5 | 83,1 | 59,7 | 69,6 |
| *09 Gotlands län* | n.a. | 38,7 | 80,2 | 38,7 | 80,2 | 61,2 | 61,2 |
| *10 Blekinge län* | 90,8 | 44,5 | 79,2 | 64,0 | 82,9 | 62,2 | 68,4 |
| *12 Skåne län* | 85,9 | 44,4 | 83,7 | 69,7 | 83,7 | 65,9 | 72,7 |
| *13 Hallands län* | 93,2 | 43,1 | 81,1 | 65,4 | 84,6 | 63,2 | 70,0 |
| *14 Västra Götalands län* | 81,2 | 44,6 | 82,9 | 67,8 | 80,4 | 65,0 | 70,4 |
| *17 Värmlands län* | 88,2 | 39,9 | 74,3 | 66,5 | 79,7 | 57,0 | 66,5 |
| *18 Örebro län* | 77,1 | 46,8 | 82,2 | 62,7 | 79,1 | 65,1 | 68,0 |
| *19 Västmanlands län* | 73,3 | 49,0 | 83,9 | 61,3 | 78,5 | 67,2 | 68,1 |
| *20 Dalarnas län* | 94,2 | 38,9 | 73,9 | 69,3 | 82,7 | 56,6 | 68,6 |
| *21 Gävleborgs län* | 90,8 | 43,3 | 78,7 | 67,4 | 83,1 | 61,3 | 69,5 |
| *22 Västernorrlands län* | 87,4 | 41,9 | 76,8 | 63,2 | 80,3 | 59,5 | 66,4 |
| *23 Jämtlands län* | 89,0 | 37,1 | 74,5 | 62,2 | 79,5 | 56,5 | 64,7 |
| *24 Västerbottens län* | 74,5 | 38,6 | 70,9 | 58,3 | 71,7 | 54,3 | 60,2 |
| *25 Norrbottens län* | 88,8 | 41,9 | 74,6 | 67,4 | 80,6 | 57,8 | 67,4 |

Column (a) of Table 3 teaches us, for example, that the economic activity indicated as $H_G$ / max($H_G$) is most centralized in Stockholm (67.9%) and most decentralized in Dalarne (94.2%). This means that the number of firms in Dalarne is distributed among the municipalities in the county more equally than in Stockholm. Gotlands län, however, contains only a single municipality at the NUTS 5 level, and therefore no uncertainty can be specified in the geographical distribution of this county.

Uncertainty in the distribution of firm-sizes is specified in column (b) of Table 3. $H_O$ is lowest in Gotland (38.7%) and highest in Västermanlands län (49.0%). One can expect these uncertainties to be relatively low because of the skew in the distributions (see Table 2 above). A $H_O / max(H_O)$ of 38.7% in Gotland indicates the relatively small number of larger firms in this country, in contrast to Västermanlands län (49.0%) where the numbers of firms of different sizes are more equally distributed. The highest level of sectorial diversification, indicated by $H_T / max(H_T)$, is found in Skåne (83.7%) and Västermanlands län (83.9%). The lowest value of this parameter is found for Västerbottens län (70.9%), indicating a relatively more specialized industry structure.

The combined uncertainties (that is, joint entropies) in two dimensions ($H_{GT}$, $H_{GO}$, $H_{TO}$) reduce uncertainty at the systems level because of mutual information (co-variation) between the dimensions (Equation 1). $H_{GO}$ is highest in Skåne (69.7%) and Dalarna län (69.3%), indicating that firms of all sizes are distributed across municipalities in these counties. The lowest value for this parameter is found in Uppsala län (56.0%). (The low value for Gotland should be disregarded because, as noted, uncertainty in the geographical distribution cannot be specified for this county.) $H_{GT}$ is highest in Sødermanlands and Hallands län (84.6%) and lowest in Västrebotten (71.7%). A high value on this parameter suggests a weaker linkage between geography and technology; companies in various industry sectors are thus more distributed. However, all these differences among regions remain moderate. $H_{TO}$ is highest in Västermannsland (67.2%), Skåne (65.9%), and Ostergøtalands län (65.9%). These three counties have Sweden's highest level of possible combinations between technological and organizational classes. The lowest value for this indicator is again found in Västrebotten (54.3%).

**Table 4**: Normalized mutual information values in two and three dimensions; NUTS3 level; $N = 1,187,142$.

|  | $\Delta T_{GO}$ | $\Delta T_{GT}$ | $\Delta T_{TO}$ | $\Delta T_{GTO}$ | N of firms |
|---|---|---|---|---|---|
| 01 Stockholms län | 1.25 | 14.66 | 21.86 | -3.49 | 281,786 |
| 03 Uppsala län | 0.17 | 2.58 | 4.39 | -0.51 | 39,664 |
| 04 Södermanlands län | 0.29 | 0.68 | 3.43 | -0.37 | 27,357 |
| 05 Östergötlands län | 0.38 | 2.50 | 5.77 | -0.59 | 43,369 |
| 06 Jönköpings län | 0.32 | 2.17 | 8.66 | -0.73 | 42,142 |
| 07 Kronobergs län | 0.18 | 1.09 | 5.22 | -0.36 | 27,267 |
| 08 Kalmar län | 0.34 | 1.52 | 5.14 | -0.72 | 30,779 |
| 09 Gotlands län | 0.00 | 0.00 | 1.09 | 0.00 | 9,570 |
| 10 Blekinge län | 0.03 | 0.27 | 2.75 | -0.33 | 16,205 |
| 12 Skåne län | 1.38 | 13.23 | 14.32 | -2.31 | 145,065 |
| 13 Hallands län | 0.28 | 2.02 | 5.34 | -0.35 | 39,232 |
| 14 Västra Götalands län | 2.13 | 25.19 | 24.52 | -2.91 | 188,819 |
| 17 Värmlands län | 0.69 | 3.14 | 7.38 | -0.49 | 38,241 |
| 18 Örebro län | 0.28 | 1.61 | 4.39 | -0.62 | 29,106 |
| 19 Västmanlands län | 0.19 | 1.48 | 3.52 | -0.49 | 24,563 |
| 20 Dalarnas län | 0.50 | 1.96 | 7.32 | -0.73 | 41,644 |
| 21 Gävleborgs län | 0.34 | 1.71 | 5.72 | -0.41 | 33,771 |
| 22 Västernorrlands län | 0.15 | 1.13 | 5.87 | -0.37 | 31,922 |
| 23 Jämtlands län | 0.23 | 1.29 | 3.90 | -0.40 | 24,060 |
| 24 Västerbottens län | 0.28 | 2.34 | 7.94 | -0.93 | 39,028 |
| 25 Norrbottens län | 0.21 | 1.34 | 7.02 | -0.83 | 33,831 |
| Sum | 9.63 | 81.92 | 155.57 | - 17.95 | 1,187,421 |
| $T_0$ | 3.34 | 93.54 | -4.43 | -4.61 |  |
| Sweden | 12.96 | 175.45 | 151.12 | -22.56 | 1,187,421 |

Table 4 shows the values of the mutual information in two and three dimensions; all values are normalized for the number of firms ($\Delta T_i = n_i T_i/N$; see Equation 2 above), and can therefore be added up and compared. A low value of mutual information—or co-variation—between the distributions in the geography and industrial sectors is found in the counties in the environments of Stockholm (Södermanland; $T_{GT}$ = 0.68 mbits) and Malmø (Blekinge; $T_{GT}$ = 0.27 mbits), indicating a diversified industry structure, as might be expected in the neighborhood of large cities (Lengyel & Leydesdorff, 2011). The region surrounding Gothenburg (Västre Götaland), however, includes a large number of smaller municipalities (49). This adds to the uncertainty in this interaction. Note that Västre Gøtalands län (25.19 mbits) shows values even higher than Stockholm (14.66 mbits) on this mutual information. The industrial and geographical clustering in this region is more specialized than in the Stockholm area.

The (normalized!) mutual information between technology and organization is always larger than the respective couplings of technologies or organizational formats to the geographical distributions. Whereas $T_{GO}$ and $T_{GT}$ can be considered as indicators of geographical clustering, $T_{TO}$ can also be considered as a correlation between the maturity of the industry and the size of the firms involved (Leydesdorff *et al.,* 2006). The highest values for this parameter are found for the three metropolitan areas (Stockholm: 21.86 mbits; Gothenburg [Västre Gøtalands län]: 24.52 mbits; and Skåne: 14.32 mbits); and the lowest values for the environments of these large cities (Södermand län around Stockholm: 3.43 mbits; Blekinge län in the environment of Malmö: 2.75 mbits). The metropoles in this case seem to have an effect upon the regions surrounding them. The national level does not add to the uncertainty in this coupling more than the sum of the regions ($T_0 < 0$; see footnote 7).

The focus of this study is on $\Delta T_{GTO}$: the normalized contribution to the mutual information in three dimensions of the regions and the nation. As expected, this parameter is highest for Stockholm (–3.49 mbits), Västre Gotalands län (–2.91 mbits), and Skåne (–2.31 mbits). In accordance with the expectation, these three counties dominate the picture of the synergy within the nation; together, they account for (8.71 / 17.95 =) 48.5% of the summed synergies of regions at the geographical scale of NUTS3. The contribution ($\Delta T_{GTO}$) to the reduction of uncertainty in three dimensions in the 21 counties is overlaid on the map of Sweden in Figure 3.

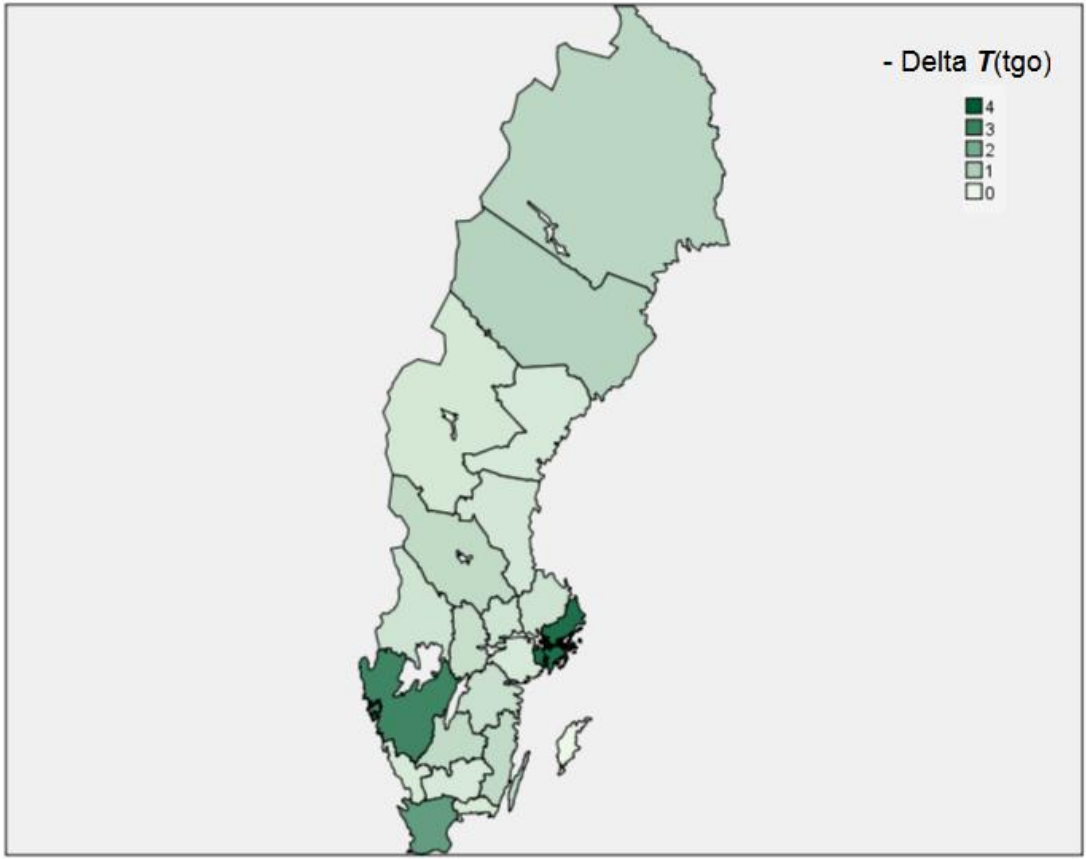

**Figure 3**: Contributions to the reduction of uncertainty at the level of 21 Swedish counties.

Of the three metropolitan areas, the capital region of Stockholm is by far the most concentrated in terms of economic activity (lowest value of $H_G/\max(H_G)$ in Table 3) and mutual information between the geographical distribution of firms and their organizational size (lowest value of $\Delta T_{GO}$ in Table 4). The uncertainty in the joint distribution of geographical addresses and technological sectors is also lower than for the two other counties; this applies similarly for $H_{GTO}$.

All these distributions are more skewed in Stockholm than in Gothenburg and Malmö. Most importantly, the synergy function $\Delta T_{GTO}$ is far more negative for Stockholm. Thus, this highly pronounced profile operates in terms of the functions of wealth generation, innovation, and control. In summary, the different functions are more concentrated and the distributions operate more synergistically in the Stockholm region than around Gothenburg and Malmö/Lund. The other regions follow these three regions at a considerable distance.

*b. Results at geographical scales above NUTS3*

In addition to being divided into 21 counties (NUTS3), Sweden is administratively organized at the NUTS2 level into eight national areas (*Riksområden*) and three *Landsdelar* at the NUTS1 level. The data allow us to aggregate at these levels and specify between-group uncertainty (Equation 2). The hierarchy among the geographical levels offers us a possibility to test whether the knowledge base of the Swedish economy is nested accordingly (Table 5).

Table 5. Between-group synergy at different geographical scales.

| Geographical scale | $\Sigma T$ in mbits | $T_0$ | $T_0$ as % contribution |
|---|---|---|---|
| NUTS0 (national level) | -22.56 | | |
| NUTS1 (3 Landsdelar) | -22.08 | -0.48 | 2.2 |
| NUTS2 (8 Riksområden) | -19.84 | -2.72 | 13.7 |
| NUTS3 (21 Counties) | -17.95 | -4.61 | 25.7 |

Using the same results as in Table 4 above, Table 5 shows first that the surplus of the national system (on top of the aggregation of the results at individual counties) is –4.61 mbit. This is 25.7% of the −22.56 mbit measured for Sweden as a national system. In other words, one quarter of the reduction of uncertainty in the national system is realized at levels higher than the regions (at NUTS3-county level). The regions, in other words, have only a moderate function in the knowledge-based economy of Sweden in comparison with other nations.

At the next level of aggregation (NUTS2), an additional reduction of uncertainty of (22.6–19.8) = 2.8 mbits, that is, 13.7%, is realized. Among the three *Landsdelar* (NUTS1), however, only 0.5 mbit, that is, 2.2% of the national sum total, is reduced by this further aggregation. In other words, the Swedish national system is organized hierarchically, as suggested by most of the literature about Sweden. The distinction between geographical units at the NUTS2 level does not add much to the regional differentiation. As noted above, this level is not in use by the Swedish administration.

How do these results relate to other indicators of the knowledge-based economy? Eurostat (2011: 204) shows that eight of the most R&D-intensive regions at the NUTS2 level are located in the Nordic countries, and of these four are in Sweden. These regions are: Sydsverige (4.75%), Västsverige (3.72%), Östra Mellansverige (3.74%), and Stockholm (4.03%). Our indicator shows that these four regions also have the highest values at the NUTS2 level: Stockholm (-3.49 mbits), Västsverige (-3.44 mbits), Östra Mellansverige (-3.21 mbits), and Sydsverige (-2.88 mbits).

The percentage of researchers as part of the population is sometimes used as another indicator for the knowledge base of an economy. From this perspective, the largest percentage is found in Stockholm and Östra Mellansverige. Eurostat indicators for EPO patents at the NUTS3 level (2006 data) show the largest numbers in the following regions: Skåne, Västre Götaland, Stockholm, Uppsala, and Västermannsland. At this geographical scale (NUTS3), our Triple Helix indicator points to the first three of these regions: Stockholm, Västre Götaland and Skåne (see Table 4).

In summary, our results are consistent (*i*) for different geographical scales and (*ii*) with other indicators of the knowledge-based economy. However, information theory enables us to specify quantitatively how much synergy is generated and at which level. The three regions in Sweden around Stockholm, Gothenburg, and Malmö together provide 48.5% of the knowledge base at the regional level. In addition to confirming the expectation, these results provide us with some confidence in the much more debatable results that we found for Norway, and earlier for Hungary.

As in the case study of the Netherlands, the Triple Helix indicator provides us with the conclusion that Sweden can be considered as a national system of innovations. In the case of Germany, in our opinion, the main result from the perspective of economic geography was that the East-West divide is no longer evident at lower levels of aggregation (NUTS2; *Regierungsbezirke*) while this divide still prevails at the level of the federal states (NUTS1; *Länder*). In the case of Hungary, a surplus value at the national level could not be found.

**Conclusions and discussion**

Swedish data provide us with a test case for the Triple Helix indicator. Both the theoretical discussions (Krippendorff, 2009a and b; Leydesdorff, 2009, 2010b) and our sometimes counter-intuitive results (Lengyel & Leydesdorff, 2010; Strand & Leydesdorff, in press) had made it urgent to find a yardstick for the measurement: can reduction of uncertainty be used as a measure of the knowledge base of an economy, and how precise is this operationalization? The results for Sweden provide us with some confidence in both the operationalization and the precision of the measurement.

Geographical decomposition is one among the possible decompositions. A complex dynamics can be expected to contain one more degree of freedom than the networks thus decomposed. Elsewhere, we have decomposed in terms of the sectors. The Swedish sector codes do not correspond precisely to the NACE codes of the OECD, so that the decomposition cannot be made unambiguously in this dimension.[8] However, from the previous studies we obtained as conclusions that knowledge-intensive services tend not to contribute to a regional economy because service providers can easily travel across regions. This mobility is less prominent for high-tech services which may require localized facilities such as laboratories and mainframe computers. High-tech manufacturing, however, is less "embedded" than medium-tech manufacturing. In other words, medium-tech manufacturing can be considered as the backbone of the knowledge-based geography (cf. Cohen & Levinthal, 1990).

---

[8] More recently, we have obtained Swedish data with NACE codes at the two-digit level. This data enables us to integrate the Swedish and Norwegian data into a single design. We intend to use this data in a more in-depth comparison of the Norwegian and Swedish innovation systems (Strand & Leydesdorff, in preparation).

Comparing the Swedish case with the previous case studies, our results suggest that both the Netherlands and Sweden function as national innovation systems. Among our case studies, these two countries have known a national system in a relatively continuous history since Napoleonic times. Germany is the most complex case in Europe, whereas Hungary went through a transition at the time of the emergence of the knowledge-based economies. Knowledge-based economies became increasingly important during the 1980s as the discussion faded about organizing political economies between liberal democracies, on the one hand, and communist states on the other (Leydesdorff, 2006: 206 ff.). When Hungary went "into transition" and "accessed" the European Union after 1990, it was too late to establish a nation state without at the same time developing a structure of the knowledge-based economy functioning at the next-order (European) level.

Norway is not part of the EU, although this country is deeply interwoven culturally with other Nordic countries such as Sweden and Denmark. One major factor in the Norwegian economy is the marine and maritime industries that have flourished since oil and gas were found offshore in the 1960s (Fagerberg *et al.*, 2009a and b). The offshore industry is highly knowledge-intensive, but also required the generation of new medium-tech companies along the west coast of the country. The universities, however, were established in a previous period, and are located in traditional centers of national culture such as Oslo and Trondheim. Thus, the different functions in the knowledge-based economy have been differentiated across various regions. In the Norwegian case, the national system of innovations adds surplus value to the reduction of uncertainty more than the sum of the parts, but this surplus at the national level is an order of magnitude smaller when compared with Sweden.

The discussion thus brings us back to the difference between the institutional and the functional interpretation of the Triple Helix model. In the institutional version, the focus is on local integration into university-industry-government networks using neo-institutional theorizing about arrangements which may construct competitive advantages. One can analyze networks using social network analysis in terms of densities, shortest distances, and brokerage functions. In the neo-evolutionary interpretation of the Triple Helix, institutions are considered as the carriers of a complex dynamics that can shape windfalls in which uncertainty is reduced.

The three core functions in a knowledge-based economy are: economic exchange, novelty production, and regulative control. Resonances between two of the three functions may lead to trajectory formation and lock-in, as when the market is lamed because of too much state control (e.g., in the energy or health sector), or when the knowledge dynamics is not sufficiently set free as in the former Soviet Union. In a complex dynamics, the three subdynamics operate upon one another, and thus upset previously constructed quasi-equilibria, leading the systems into new regimes. Such a regime transition happened for example in Hungary, and in different forms also in Germany and Norway. The German system was able to absorb the GDR at lower levels of aggregation during the transition. Norway developed into a knowledge-based economy which is integrated not nationally, but globally.

Systems such as Sweden and Holland, which because of contextual stabilities were able to make the transition to a knowledge-based economy more smoothly during the '80s and '90s, may not sufficiently open up to the challenges and become "locked-in" into the institutional arrangements of the previous period. In Sweden, the innovative agencies have been acutely aware of these problems, and Triple Helix arrangements were most emphatically stimulated during the early 2000s (Etzkowitz, 2005). The Swedish agency for innovation, VINNOVA, championed the "Triple Helix" model and, in accordance with the country's neo-corporatist traditions (Lehmbruch & Schmitter, 1982), institutional collaboration across institutional spheres increasingly became a condition for research funding in Sweden. However, a neo-evolutionary perspective makes us aware that this is a specific strategy which leads to more stability in a system than may be desirable in current international environments.

In another context, for example, Ye *et al*. (in preparation) note that a national system can also be too integrated locally in terms of Triple Helix relations in this age of globalization. Although comparable on various dimensions, Brazil and Indonesia, for example, exemplify nations where functional differentiation and institutional integration across spheres have been shaped differently. One national model cannot be transferred to another country, but modeling and measurement can make us aware of how the differences between local integration at the institutional level can be appreciated as a trade-off against openness to differentiation at the global level, and this trade-off can then perhaps be made the subject of future innovation policies.

**Appendix A**. Firm classification codes in use by Statics Sweden (2011a).[9]

| Code | Description SNI 2007 |
|------|----------------------|
| 0 | Unclassified |
| A | Agriculture, Forestry and Fishing |
| B | Mining and Quarrying |
| C | Manufacturing |
| D | Electricity, Gas, Steam and Air Conditioning Supply |
| E | Water Supply; Sewerage, Waste Management and Remediation Activities |
| F | Construction |
| G | Wholesale and Retail Trade; Repair of Motor Vehicles and Motorcycles |
| H | Transportation and Storage |
| I | Accommodation and Food Service Activities |
| J | Information and Communication |
| K | Financial and Insurance Activities |
| L | Real Estate Activities |
| M | Professional, Scientific and Technical Activities |
| N | Administrative and Support Service Activities |
| O | Public Administration and Defence; Compulsory Social Security |
| P | Education |
| Q | Human Health and Social Work Activities |
| R | Arts, Entertainment and Recreation |
| S | Other Service Activities |
| T | Activities of Households as Employers; Undifferentiated Goods- and Services-Producing Activities of Households for Own Use |
| U | Activities of Extraterritorial Organisations and Bodies |

---

[9] These codes are less fine-grained than the NACE codes of the OECD. A concordance table can be found at http://www.scb.se/Grupp/Hitta_statistik/Forsta_Statistik/Klassifikationer/_Dokument/070129kortversionSnisorterad2007.pdf .